\documentclass{PoS}
\usepackage[authoryear]{natbib}
\usepackage{graphicx}
\usepackage{amsmath}
\usepackage{amssymb}
\usepackage{txfonts}

\title{Monitoring of pulse period in Her X-1 with \textsl{Swift/BAT}:
evidence of mass ejection}

\ShortTitle{Monitoring of pulse period in Her X-1}

\author{\speaker{D. Klochkov}$^{,a}$, R. Staubert$^a$,
K. Postnov$^b$, N. Shakura$^b$, A. Santangelo$^a$\\
  \llap{$^a$}Institut f\"ur Astronomie und Astrophysik, Universit\"at
    T\"ubingen (IAAT)\\
  \llap{$^b$}Sternberg Astronomical Institute, Moscow University\\
  E-mail: \email{klochkov@astro.uni-tuebingen.de}
}

\abstract{
Monitoring of pulse period variations in accreting binary pulsars is an
important tool to study the interaction between the magnetosphere of
the neutron star and the accretion disk. While the X-ray flux
of the brightest X-ray pulsars have been successfully monitored over many
years (e.g. with RXTE/ASM, CGRO/BATSE, Swift/BAT), the possibility to
monitor their pulse timing properties continuously has so far been
very limited. In our work we use Swift/BAT observations
to study one of the most enigmatic X-ray pulsars, Hercules X-1.
For the first time, a quasi-continuous monitoring of the pulse period and
the pulse period derivative of Her X-1, is achieved over a long time (> 4\,yrs).
We argue that together with the long-term decrease of the
orbital period in Her X-1 the measured pulse period behaviour
requires the presence of mass ejection from the inner parts of the
accretion disk along the open magnetic field lines. The mass ejection
episodes probably take place during strong spin-down episodes which are
associated with the low X-ray luminosity.
}

\FullConference{The Extreme sky: Sampling the Universe above 10 keV - extremesky2009,\\
		October 13-17, 2009\\
		Otranto (Lecce) Italy}

\begin{document}

\section{Introduction}

The persistent accreting pulsar Hercules X-1 was one of the first
X-ray sources discovered by the \textsl{Uhuru} satellite in 1972
\citep{Tananbaum_etal72,Giacconi_etal73} and since then it remains
one of the most intensively studied X-ray pulsars. The mass of the 
optical companion is $\sim$2$M_{\odot}$ 
which places the system in the middle between high and low mass X-ray 
binaries. Other main parameters of the binary system are the following:
orbital period $P_\text{orb}\simeq 1.7$\,days, 
spin period of the neutron star $P_\text{spin}\simeq 1.24$\,sec, 
X-ray luminosity of the 
source $L_{\rm X}\sim 2\times 10^{37} {\rm erg\,s}^{-1}$ for a distance of 
$\sim$7~kpc \citep{Reynolds_etal97}. The binary orbit is almost 
circular \citep{Staubert_etal09}
and has an inclination $i\sim 85-88^{\circ}$ \citep{GerendBoynton76}.

Like many other X-ray pulsars Her X-1 exhibits significant 
variation of the pulsation period.
Alternation of spin-up and spin-down episodes on time-scales of several
months in this system is superimposed on a background of systematic spin-up 
\citep[e.g.][]{Staubert_etal06,Klochkov07}. The behavior of the
pulsar's spin period on shorter time scales is not very well studied because
such a study would require a continuous monitoring of \hbox{Her X-1} with a
sensitive X-ray detector. Only between 1991 and 2000 the \textsl{BATSE} 
instrument onboard \textsl{CGRO} 
provided the information about the source's pulse period on a regular basis. 

In this work we present a continuous monitoring of the \hbox{Her X-1} pulse
period $P$ and its local (measured at the time of the observation) 
time derivative $\dot P$ using the \textsl{Swift/BAT} instrument
starting from 2005 (begin of scientific operation) to 2009. 
The data of the \hbox{monitoring} allowed us to explore for the first time the
correlation between the {\em locally} measured $\dot P$ and the X-ray flux
of the pulsar and compare the results with predictions of the accretion
theory. The observed strong spin-down episodes are discussed in the frame 
of a model assuming ejection of matter from the inner part of the 
accretion disk along the open magnetic field lines.

\section{Observations and data analysis}

For our analysis we used the public archival data obtained with the 
\textbf{B}urst \textbf{A}lert \textbf{T}elescope 
(\textsl{BAT}, 150--150\,keV, \citealt{Barthelmy_etal06}) 
onboard the \textsl{Swift} observatory \citep{Gehrels_etal04}.
With its large field of view (1.4 sterad) the \textsl{BAT} instrument 
is originally designed to provide fast triggers for gamma-ray bursts 
and their accurate positions in the sky ($\sim$4\,arcsec). 
Following such a trigger, the observatory points in the direction of the burst,
which can be then observed with the X-ray and UV/optical telescopes
onboard the satellite. While searching for bursts, \textsl{BAT} 
points at different locations of the sky, thus, performing 
an all-sky monitoring in hard X-rays (measurements of the X-ray flux
are provided by the \textsl{Swift/BAT} team in the form of X-ray light curves
for the several hundred bright persistent and transient 
sources\footnote{http://heasarc.gsfc.nasa.gov/docs/swift/results/transients/}).
If a bright pulsating source
with a known period falls into the field of view of the instrument, 
the total count rate stored by \textsl{BAT} can be used to search 
for coherent pulsations of 
that source. We have used this strategy to measure the 1.24\,s pulsations
of \hbox{Her X-1} during its so-called {\em main-on} states 
(when the X-ray flux of the source is high). Such states repeat 
every $\sim$35 days and are believed to be caused by a precessing tilted 
accretion disk around the neutron star, see e.g. \citealt{GerendBoynton76}.

To determine the pulse period, we used the method similar to that
described by \citet{Staubert_etal09} which includes two techniques 
for the determination of the period: \textsl{epoch folding} 
with $\chi^2$ search 
\citep[e.g.][]{Leahy_etal83} and \textsl{pulse phase connection}
\citep[e.g.][]{Deeter_etal81}. The first one is used to establish the
presence of the periodic signal from \hbox{Her X-1} in the \textsl{BAT} data,
determine the approximate period, and construct pulse profiles 
(by folding the data with the found period), while the second 
is subsequently applied to the pulse profiles to determine the
precise value of the period and its time derivative.

For our analysis we used the total count rates 
measured by \textsl{BAT} with a time resolution of 64 msec. All times 
of the count rates were translated to the solar system barycenter and 
corrected for binary motion (using our newest orbital ephemeris presented in
\citealt{Staubert_etal09}). Then we performed a period search using
\textsl{epoch folding} in a narrow period interval around the expected
pulse period ($\sim$1.237\,s). If a strong periodic signal is present
we determined the period and used it to construct X-ray pulse profiles for 
subsequent \textsl{pulse phase connection}. 

\begin{figure}
  \begin{minipage}{0.6\textwidth}
    \resizebox{\hsize}{!}{\includegraphics{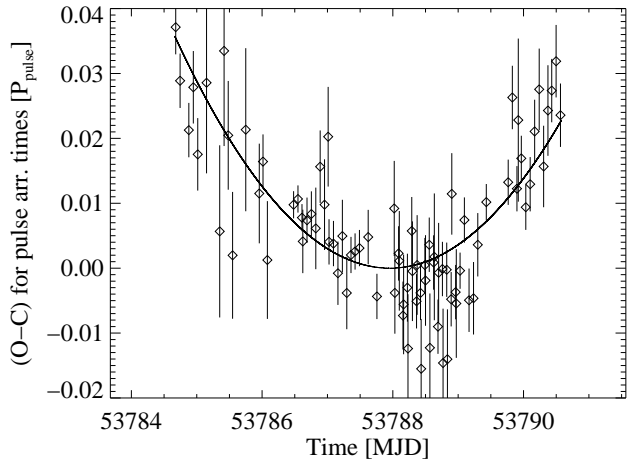}}
  \end{minipage}
  \begin{minipage}{0.35\textwidth}
    \caption{
    Estimated (observed) minus calculated (assuming a constant period) 
    pulse arrival times of \hbox{Her X-1} in units of its pulse period 
    during one of its main-on states observed with \textsl{Swift/BAT}.
    Parabolic fit to the data shown with the solid line corresponds to
    a constant positive $\dot P\simeq 1.4\times 10^{-12}$\,s/s. \label{oc}}
  \end{minipage}
\end{figure}

A convenient way to explore the variation of the pulse period,
often used in phase-connection technique, 
is to construct the so-called $(O-C)$ diagram
showing the estimated (observed) pulse arrival time minus the
calculated one assuming a constant period. An example of such a diagram
measured with \textsl{BAT} during one of the \hbox{Her X-1} 
main-on states is shown in Fig.\,\ref{oc}. A straight line in the
graph would correspond to a constant period defined by the slope
of the line. The solid curve indicates
the parabolic fit to the data corresponding to a constant
positive $\dot P$. 

\section{Results}

Using the method described in the previous section we determined the
pulse period $P$ and its time derivative $\dot P$ for most of the 
\hbox{Her X-1} main-on states observed with \textsl{BAT}.
For other parts of the 35\,d cycle, i.e. outside main-ons,
the flux was too low for such determinations.
At the time of preparing this text the data are available for the time period
from March 2005 to May 2009 that covers 45 35d cycles of the pulsar. 
For several cycles the \textsl{BAT} observations 
have relatively poor statistics due to gaps in the data. For such cycles
only $P$, but no  $\dot P$ could be determined.

\begin{figure}
  \begin{minipage}{0.6\textwidth}
    \resizebox{\hsize}{!}{\includegraphics{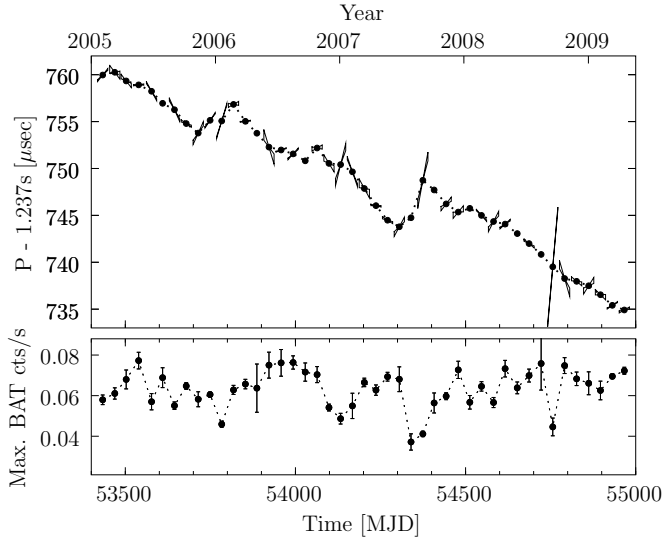}}
  \end{minipage}
  \begin{minipage}{0.35\textwidth}
    \caption{
      \textbf{Top:}
      Pulse period $P$ of Her X-1 measured with \textsl{Swift/BAT} as 
      a function of time. The cones around each point indicate the 
      allowed range of the slope corresponding
      to the measured $\dot P$ and its uncertainties. Measurement errors
      of the period itself are smaller than the symbol sizes.
      \textbf{Bottom:} Maximum \textsl{Swift/BAT} count/rate in the respective
      main-on state.
      \label{per}}
  \end{minipage}
\end{figure}

The time evolution of the measured pulse period of
\hbox{Her X-1} is shown in Fig.\,\ref{per} together with the X-ray flux in
the respective main-on states. Where a corresponding
value of $\dot P$ was measured, the 1$\sigma$-uncertainty range is indicated
by the cones, the orientation 
of which reproduce the measured $\dot P$ value.

One can see, that for many points the locally measured $\dot P$  
is substantially different from the one derived from $P$ values of 
adjacent measurements (that is from the slope of the pulse period
development). Thus, one can conclude that strong pulse period 
variations in \hbox{Her X-1} occur on shorter time scales than the 35\,d 
super-orbital period of the system.

\begin{figure}
  \begin{minipage}{0.6\textwidth}
    \resizebox{\hsize}{!}{\includegraphics{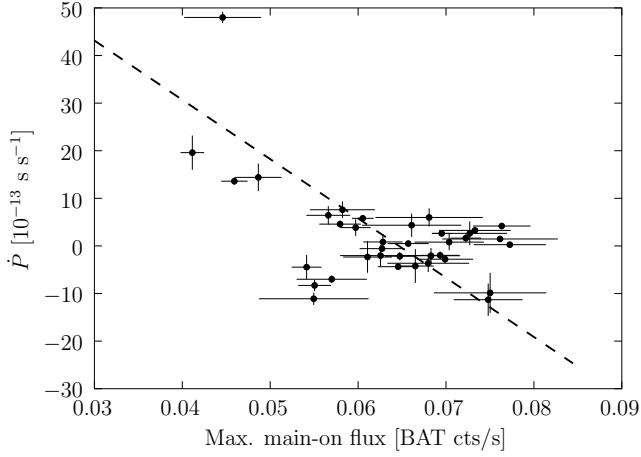}}
  \end{minipage}
  \begin{minipage}{0.35\textwidth}
    \caption{
      The locally measured time derivative of the pulse period, $\dot P$,
      in \hbox{Her X-1} versus the maximum X-ray flux during the main-on state
      as determined from the \textsl{Swift/BAT} data. A linear fit (taking
      uncertainties of both variables into account) is shown with the dashed 
      line. \label{cor}}
  \end{minipage}
\end{figure}

Already in Fig.\,\ref{per} one can notice that the strong spin-down episodes
occasionally exhibited by Her X-1 and lasting from 1 to 3 main-ons are
mostly coincident with drops in the X-ray flux. This behavior can also
be demonstrated by plotting values of $\dot P$ versus X-ray flux
(Fig.\,\ref{cor}). The data indicate an 
anticorrelation as predicted by the basic accretion theory. Inspection
of the linear Pearson's correlation coefficient gives the probability
of $\sim$4$\times 10^{-4}$ to find the measured correlation by chance.

The correlation is mainly driven by the group of four points with high 
spin-down rate and low  flux (in the upper left part of the graph in 
Fig.\,\ref{cor}). The rest of the points forms an uncorrelated "cloud" 
around $\dot P = 0$. This means that most of the time the pulsar is close
to the equilibrium regime, when spin-up and spin-down torques are nearly
balanced. Sometimes, however, it switches to strong spin-down accompanied
by strong decrease of the X-ray flux. Such a behavior needs further 
explanation. In the next section we argue that the episodes of strong
spin-down are associated with strong matter outflow from the inner parts
of the accretion disk. The mass transfer rate through the accretion disk
does not need to change significantly.

\section{Evidence for the coronal mass ejection\label{eject:discussion}}

If one assumes that the accretion disk carries some
magnetic field, it can interact via reconnection with the neutron 
star's magnetosphere beyond the corotation radius $R_c$. This might imply 
that beyond  $R_c$ the field lines can sometimes inflate to become open 
\citep[see also discussion in][]{Lovelace_etal95}.
During such episodes, a substantial fraction of matter in the inner part of
the accretion disk can escape the system in the form of a coronal 
wind ejection along the open field lines. 

Such an ejection of matter should be reflected 
in a secular change of the system's
orbital period which is indeed observed in \hbox{Her X-1} 
\citep{Deeter_etal91,Staubert_etal09}.
To assess the importance of coronal mass ejections for the orbital period 
evolution, we invoked general considerations of the non-conservative 
treatment of binary orbital parameters \citep[see e.g.][]{Grishchuk_etal01}.
First, we defined the non-conservativeness parameter in the 
standard way \citep[e.g.][]{RitterKolb92}: 
$\eta=-{\dot M_x}/{\dot M_o}\le 1$, ($\dot M_o<0$).
Assume that the ejected mass carries away the specific angular 
momentum of the neutron star 
we express the fractional change of the orbital period through 
$\dot M_x/M_x$, $q$ and $\eta$:
\begin{equation}
\frac{1}{3}\frac{\dot P_{\rm orb}}{P_{\rm orb}}=-\frac{\dot M_x}{M_x}\left[
1-\frac{q}{\eta}-\left(1-\frac{1}{\eta}\right)\frac{q/3+1}{q+1}
\right]\,,
\label{dPP}
\end{equation}
where $M_x$ is the mass of the accretor, $\dot M_x$ is the accretion rate
onto the neutron star, $q=M_x/M_{\rm opt}$ is the binary mass ratio.

It is convenient to normalize the mass accretion rate onto 
the neutron star $\dot M_x$  to the value that can be derived from the 
observed fractional change of the orbital period in the conservative case
$(\dot M_x)_{\rm cons}$,
so that $\dot m\equiv \dot M_x/(\dot M_x)_{\rm cons}$. 
In \hbox{Her X-1} $q\simeq 0.63$ and 
one finds $(\dot M_x)_{\rm cons}\simeq 8 \times 10^{17}$\,g\,s$^{-1}$ 
for the measured $\dot P_{\rm orb} = -4.85\times 10^{-11}$\,s\,s$^{-1}$
\citep{Staubert_etal09}.
Then we can eliminate $\dot P_{\rm orb}/P_{\rm orb}$ from the left hand side of 
Eq.\,(\ref{dPP}) to obtain the equation for $\eta$ at a given $\dot m$:
\begin{equation}
\eta=\frac{q^2+2q/3-1}{(q^2-1)/\dot m+2q/3}\,.
\label{eta}
\end{equation}
From here we see that $\dot m<1$ leads to $\eta<1$, i. e. if one wants
to decrease the mass accretion rate onto the neutron star to get 
smaller X-ray luminosity (as is the case of \hbox{Her X-1}, where the mean 
observed $L_x\sim 2\times 10^{37}$\,erg\,s$^{-1}$ is 4 times
smaller than the one following from the conservative mass 
exchange analysis, \citealt{Staubert_etal09}), 
a non-conservative mass exchange is required. Specifically,
if we want to bring in accordance the observed $\dot P_{\rm orb}$ 
and $L_{\rm X}$ in 
\hbox{Her X-1}, we would need $\dot m\simeq 1/4$ and (from Eq.\,\ref{eta}) 
$\eta\sim 0.1$, a fairly high non-conservative mass exchange.

In the frame of our model, the mass ejection from the system
through the open magnetic field lines occurs most efficiently
during strong spin-down episodes which are associated with small X-ray 
luminosity.  
Indeed, as we see in Fig\,\ref{cor}, the observed 
X-ray flux is decreased by a factor of two during strong spin-down. 
From Eq.\,(\ref{eta}) it is easy to find that
at a given $q$ a fractional decrease in $\dot m$ leads to comparable 
fractional decrease in $\eta$, i.e. accretion indeed becomes more 
non-conservative during strong mass ejection episodes.
During such episodes, the neutron star spin-down power
$I\omega \dot \omega$ is spent to expel accreting matter from the inner disk
radius $R_d\sim R_c$:
\begin{equation}
I\omega \dot \omega \sim \dot M_{\rm ej}\frac{GM}{R_c}\,.
\end{equation}
This equation is satisfied for the observed parameters of \hbox{Her X-1}: 
ejected mass rate during strong spin-downs 
$\dot M_{\rm ej}\sim 0.5\dot M_x\simeq 10^{17}$\,g\,s$^{-1}$, 
$\dot P \simeq 10^{-12}$\,s\,s$^{-1}$, and $R_c\simeq 1.3\times 10^8$\,cm.

\bibliographystyle{aa}
\bibliography{ref}

\begin{thebibliography}{15}
\expandafter\ifx\csname natexlab\endcsname\relax\def\natexlab#1{#1}\fi

\bibitem[{{Barthelmy} {et~al.}(2005){Barthelmy}, {Barbier}, {Cummings},
  {Fenimore}, {Gehrels}, {Hullinger}, {Krimm}, {Markwardt}, {Palmer},
  {Parsons}, {Sato}, {Suzuki}, {Takahashi}, {Tashiro}, \&
  {Tueller}}]{Barthelmy_etal06}
{Barthelmy}, S.~D., {Barbier}, L.~M., {Cummings}, J.~R., {et~al.} 2005, Space
  Science Reviews, 120, 143

\bibitem[{{Deeter} {et~al.}(1991){Deeter}, {Boynton}, {Miyamoto}, {Kitamoto},
  {Nagase}, \& {Kawai}}]{Deeter_etal91}
{Deeter}, J.~E., {Boynton}, P.~E., {Miyamoto}, S., {et~al.} 1991, ApJ, 383, 324

\bibitem[{{Deeter} {et~al.}(1981){Deeter}, {Boynton}, \&
  {Pravdo}}]{Deeter_etal81}
{Deeter}, J.~E., {Boynton}, P.~E., \& {Pravdo}, S.~H. 1981, ApJ, 247, 1003

\bibitem[{{Gehrels} {et~al.}(2004){Gehrels}, {Chincarini}, {Giommi}, {Mason},
  {Nousek}, {Wells}, {White}, {Barthelmy}, {Burrows}, {Cominsky}, {Hurley},
  {Marshall}, {M{\'e}sz{\'a}ros}, {Roming}, {Angelini}, {Barbier}, {Belloni},
  {Campana}, {Caraveo}, {Chester}, {Citterio}, {Cline}, {Cropper}, {Cummings},
  {Dean}, {Feigelson}, {Fenimore}, {Frail}, {Fruchter}, {Garmire}, {Gendreau},
  {Ghisellini}, {Greiner}, {Hill}, {Hunsberger}, {Krimm}, {Kulkarni}, {Kumar},
  {Lebrun}, {Lloyd-Ronning}, {Markwardt}, {Mattson}, {Mushotzky}, {Norris},
  {Osborne}, {Paczynski}, {Palmer}, {Park}, {Parsons}, {Paul}, {Rees},
  {Reynolds}, {Rhoads}, {Sasseen}, {Schaefer}, {Short}, {Smale}, {Smith},
  {Stella}, {Tagliaferri}, {Takahashi}, {Tashiro}, {Townsley}, {Tueller},
  {Turner}, {Vietri}, {Voges}, {Ward}, {Willingale}, {Zerbi}, \&
  {Zhang}}]{Gehrels_etal04}
{Gehrels}, N., {Chincarini}, G., {Giommi}, P., {et~al.} 2004, ApJ, 611, 1005

\bibitem[{{Gerend} \& {Boynton}(1976)}]{GerendBoynton76}
{Gerend}, D. \& {Boynton}, P.~E. 1976, ApJ, 209, 562

\bibitem[{{Giacconi} {et~al.}(1973){Giacconi}, {Gursky}, {Kellogg}, {Levinson},
  {Schreier}, \& {Tananbaum}}]{Giacconi_etal73}
{Giacconi}, R., {Gursky}, H., {Kellogg}, E., {et~al.} 1973, ApJ, 184, 227

\bibitem[{{Grishchuk} {et~al.}(2001){Grishchuk}, {Lipunov}, {Postnov},
  {Prokhorov}, \& {Sathyaprakash}}]{Grishchuk_etal01}
{Grishchuk}, L.~P., {Lipunov}, V.~M., {Postnov}, K.~A., {Prokhorov}, M.~E., \&
  {Sathyaprakash}, B.~S. 2001, Physics Uspekhi, 44, 1

\bibitem[{Klochkov(2007)}]{Klochkov07}
Klochkov, D. 2007, PhD thesis, University of T\"ubingen, Germany,
  http://w210.ub.uni-tuebingen.de/volltexte/2007/3181/pdf/disser.pdf

\bibitem[{{Leahy} {et~al.}(1983){Leahy}, {Elsner}, \&
  {Weisskopf}}]{Leahy_etal83}
{Leahy}, D.~A., {Elsner}, R.~F., \& {Weisskopf}, M.~C. 1983, ApJ, 272, 256

\bibitem[{{Lovelace} {et~al.}(1995){Lovelace}, {Romanova}, \&
  {Bisnovatyi-Kogan}}]{Lovelace_etal95}
{Lovelace}, R.~V.~E., {Romanova}, M.~M., \& {Bisnovatyi-Kogan}, G.~S. 1995,
  MNRAS, 275, 244

\bibitem[{{Reynolds} {et~al.}(1997){Reynolds}, {Quaintrell}, {Still}, {Roche},
  {Chakrabarty}, \& {Levine}}]{Reynolds_etal97}
{Reynolds}, A.~P., {Quaintrell}, H., {Still}, M.~D., {et~al.} 1997, MNRAS, 288,
  43

\bibitem[{{Ritter} \& {Kolb}(1992)}]{RitterKolb92}
{Ritter}, H. \& {Kolb}, U. 1992, A\&A, 259, 159

\bibitem[{{Staubert} {et~al.}(2009){Staubert}, {Klochkov}, \&
  {Wilms}}]{Staubert_etal09}
{Staubert}, R., {Klochkov}, D., \& {Wilms}, J. 2009, A\&A, [accepted], arXiv
  0904.2307

\bibitem[{{Staubert} {et~al.}(2006){Staubert}, {Schandl}, {Klochkov}, {Wilms},
  {Postnov}, \& {Shakura}}]{Staubert_etal06}
{Staubert}, R., {Schandl}, S., {Klochkov}, D., {et~al.} 2006, in American
  Institute of Physics Conference Series, Vol. 840, The Transient Milky Way: A
  Perspective for MIRAX, ed. F.~{D'Amico}, J.~{Braga}, \& R.~E. {Rothschild},
  65--70

\bibitem[{{Tananbaum} {et~al.}(1972){Tananbaum}, {Gursky}, {Kellogg},
  {Levinson}, {Schreier}, \& {Giacconi}}]{Tananbaum_etal72}
{Tananbaum}, H., {Gursky}, H., {Kellogg}, E.~M., {et~al.} 1972, ApJ, 174, L143

\end{thebibliography}

\end{document}